# Decoupling of Eulerian and Lagrangian variables in Lagrangian velocity correlations.

Moshe Schwartz

Raymond and Beverly School of Physics and Astronomy

Tel Aviv University, Tel Aviv 69978, Israel

bricki@netvision.net.il

Abstract: The motion of a particle carried by a liquid is described by the differential equation $\dot{\mathbf{r}} = \mathbf{v}(\mathbf{r},t)$, where $\mathbf{v}$ is the Eulerian velocity field of the liquid. Assuming the velocity field to be random with a stationary, isotropic, translational invariant, zero mean distribution, the Lagrangian velocity correlation of the particle can be expressed in terms of the Eulerian correlations and the function $f(q,t) = \langle \exp i\mathbf{q} \cdot \mathbf{r}(t) \rangle$, where the particle is taken to be at the origin at time $t=0$ and $\langle \cdots \rangle$ denotes averaging over the Eulerian velocity distribution. This is a result of a decoupling, which is exact to leading order in the volume of the system.

Consider an incompressible liquid confined in a large cubical volume $\Omega$ with periodic boundary conditions. Let the Fourier coefficient of the velocity field, $\mathbf{v}(\mathbf{r},t)$, be given by

$$\mathbf{v}(\mathbf{q},t) = \frac{1}{\sqrt{\Omega}} \int_\Omega d\mathbf{r}\, \mathbf{v}(\mathbf{r},t) \exp(-i\mathbf{q} \cdot \mathbf{r}). \tag{1}$$

The velocity Fourier components of the velocity field be bounded and random with zero average and correlations,

$$\langle \mathbf{v}_i(\mathbf{q},t)\mathbf{v}_j(\mathbf{p},t') \rangle = [\delta_{ij} - \frac{\mathbf{q}_i \mathbf{q}_j}{\mathbf{q}^2}]\delta_{\mathbf{p},-\mathbf{q}}\Phi(\mathbf{q},t-t'). \tag{2}$$

Consider next a particle carried by the velocity field that obeys the equation

$$\dot{\mathbf{r}} = \mathbf{v}(\mathbf{r},t). \tag{3}$$

The problem is to obtain the Lagrangian velocity correlation of the particle,

$$\Lambda(t-t') = \langle \dot{\mathbf{r}}(t) \cdot \dot{\mathbf{r}}(t') \rangle, \tag{4}$$

where the average is over realizations of the velocity field.

It follows from equation (3) that

$$\Lambda(t-t') = \frac{1}{\Omega} \sum_{\mathbf{q},\mathbf{p}} \langle \mathbf{v}(\mathbf{q},t) \cdot \mathbf{v}(\mathbf{p},t') \exp[i(\mathbf{q} \cdot \mathbf{r}(t) + \mathbf{p} \cdot \mathbf{r}(t'))] \rangle. \tag{5}$$

Clearly, by the translational invariance of the Eulerian velocity distribution, the right hand side is unchanged if $\mathbf{r}(t)$ and $\mathbf{r}(t')$ are replaced by $\mathbf{r}(t)+\mathbf{d}$ and $\mathbf{r}(t')+\mathbf{d}$. Therefore, performing a spatial average by translating the $\mathbf{r}$ 's by $\mathbf{d}$, then integrating over $\mathbf{d}$ and dividing by the volume, $\Omega$, still leaves the right hand side unchanged. Consequently,

$$\Lambda(t-t') = \frac{1}{\Omega} \sum_{\mathbf{qp}} \langle \mathbf{v}(\mathbf{q},t) \cdot \mathbf{v}(\mathbf{p},t') \exp[i(\mathbf{q}\cdot\mathbf{r}(t)+\mathbf{p}\cdot\mathbf{r}(t'))]\rangle \delta_{\mathbf{p},-\mathbf{q}}, \quad (6)$$

where $\delta$ is the Kronecker $\delta$. Therefore,

$$\Lambda(t-t') = \frac{1}{\Omega} \sum_{\mathbf{q}} \langle \mathbf{v}(\mathbf{q},t) \cdot \mathbf{v}(-\mathbf{q},t') \exp[i(\mathbf{q}\cdot(\mathbf{r}(t)-\mathbf{r}(t')))]\rangle. \quad (7)$$

Due to invariance under time translations and taking, without loss of generality, $\mathbf{r}(0)=0$, finally

$$\Lambda(t) = \frac{1}{\Omega} \sum_{\mathbf{q}} \langle \mathbf{v}(\mathbf{q},t) \cdot \mathbf{v}(-\mathbf{q},0) \exp[i(\mathbf{q}\cdot(\mathbf{r}(t)))]\rangle. \quad (8)$$

The claim now concerns the average in the sum on the right hand side of equation (8),

$$\langle \mathbf{v}(\mathbf{q},t) \cdot \mathbf{v}(-\mathbf{q},0) \exp[i\mathbf{q}\cdot\mathbf{r}(t)]\rangle = \langle \mathbf{v}(\mathbf{q},t)\mathbf{v}(-\mathbf{q},0)\rangle \langle \exp[i\mathbf{q}\cdot\mathbf{r}(t)]\rangle \quad (9)$$

to leading order in $\Omega$.

If the claim is true then the Lagrangian correlations can be written in the limit of the infinite system as

$$\Lambda(t-t') = \frac{1}{(2\pi)^d} \int d\mathbf{q}\, [\delta_{ij} - \frac{q_i q_j}{\mathbf{q}^2}] \Phi(q,t-t') f(q,t-t'). \quad (10)$$

The decoupling given by equation (9) and the resulting equation (10) have been used successfully in the study of a number of problems: The derivation of a Langevin equation [1] and the generalized Langevin equation [2,3] from a Lagrangian describing the interaction of a particle with longitudinal phonons. The diffusion of a deformable object in a random flow [4,5] and the trajectory of a particle swept by a turbulent liquid [6]. In the following I prove that decoupling.

To prove equation (9) consider the dependence of $\exp[i\mathbf{q}\cdot\mathbf{r}(t)]$ on the Fourier components of the velocity field of the liquid with momenta $\mathbf{q}$ and $-\mathbf{q}$. Let us take first the situation, where those components are taken to be identically zero at all times. Let the corresponding solution of equation (3) with the initial condition $\mathbf{r}(0)=0$ be denoted by $\bar{\mathbf{r}}(\mathbf{q},-\mathbf{q},t)$. Equation (3) will be expressed now in the form

$$\dot{\mathbf{r}} = \frac{1}{\sqrt{\Omega}} \sum_{\mathbf{p}\neq\mathbf{q},-\mathbf{q}} \mathbf{v}(\mathbf{p},t) \exp[i\mathbf{p}\cdot\mathbf{r}(t)] + \frac{1}{\sqrt{\Omega}} \mathbf{v}(\mathbf{q},t)\exp[i\mathbf{q}\cdot\mathbf{r}(t)] + \mathbf{q}\to-\mathbf{q}. \quad (11)$$

The solution of the above may be written to order $\Omega^{-1/2}$ as

$$\mathbf{r}(t) = \bar{\mathbf{r}}(\mathbf{q},-\mathbf{q},t) + \frac{\mathbf{a}(t)}{\sqrt{\Omega}}, \tag{13}$$

where

$$\mathbf{a}(t) = \int_0^t dt' F_+\{\mathbf{v}(\mathbf{p},t'');\mathbf{q},t',t\}\mathbf{v}(\mathbf{q},t') + F_-\{\mathbf{v}(\mathbf{p},t'');-\mathbf{q},t',t\}\mathbf{v}(-\mathbf{q},t') \tag{14}$$

This is a linear functional of $\mathbf{v}(\mathbf{q},t)$ and $\mathbf{v}(-\mathbf{q},t)$ with coefficients $F_+$ and $F_-$ which are (non-linear) functionals of all the functions $\mathbf{v}(\mathbf{p},t')$ with $\mathbf{p} \neq \mathbf{q},-\mathbf{q}$.

Consider now the average on the left hand side of equation (9) and perform it in two steps. The first step involves averaging only over $\mathbf{v}(\mathbf{q},t)$ and $\mathbf{v}(-\mathbf{q},t)$. Denote this average by $\langle \cdots \rangle_{\mathbf{q},-\mathbf{q}}$. Since $\bar{\mathbf{r}}(\mathbf{q},-\mathbf{q},t)$ does not depend on $\mathbf{v}(\mathbf{q},t)$ and $\mathbf{v}(-\mathbf{q},t)$,

$$\langle \mathbf{v}(\mathbf{q},t) \cdot \mathbf{v}(-\mathbf{q},0) \exp[i\mathbf{q} \cdot \mathbf{r}(t)] \rangle_{\mathbf{q},-\mathbf{q}} = \exp[i\mathbf{q} \cdot \bar{\mathbf{r}}(\mathbf{q},-\mathbf{q},t)] \times$$

$$\times \left\langle \mathbf{v}(\mathbf{q},t) \cdot \mathbf{v}(-\mathbf{q},0) \exp \frac{1}{\sqrt{\Omega}}[i\mathbf{q} \cdot \mathbf{a}(t)] \right\rangle_{\mathbf{q},-\mathbf{q}}. \tag{15}$$

It is easy to verify that the average on the right hand side of equation (15) is given by

$$\left\langle \mathbf{v}(\mathbf{q},t) \cdot \mathbf{v}(-\mathbf{q},0) \exp \frac{1}{\sqrt{\Omega}}[i\mathbf{q} \cdot \mathbf{a}(t)] \right\rangle_{\mathbf{q},-\mathbf{q}} = \langle \mathbf{v}(\mathbf{q},t) \cdot \mathbf{v}(-\mathbf{q},0) \rangle + O(\frac{1}{\Omega}). \tag{16}$$

Consequently, after averaging over the remaining components of the velocity field

$$\langle \mathbf{v}(\mathbf{q},t) \cdot \mathbf{v}(-\mathbf{q},0) \exp[i\mathbf{q} \cdot \mathbf{r}(t)] \rangle = \langle \mathbf{v}(\mathbf{q},t) \cdot \mathbf{v}(-\mathbf{q},0) \rangle \langle \exp[i\mathbf{q} \cdot \bar{\mathbf{r}}(\mathbf{q},-\mathbf{q},t)] \rangle + O(\frac{1}{\Omega}). \tag{17}$$

Replacing in all the above considerations $\mathbf{v}(\mathbf{q},t) \cdot \mathbf{v}(-\mathbf{q},0)$ by its bound, it is easy to see that

$$\langle \exp[i\mathbf{q} \cdot \mathbf{r}(t)] \rangle = \langle \exp[i\mathbf{q} \cdot \bar{\mathbf{r}}(\mathbf{q},-\mathbf{q},t)] \rangle + O(\frac{1}{\Omega}). \tag{18}$$

Combining equations (17) and (18) the decoupling (9) is verified and its accuracy is of order $\frac{1}{\Omega}$.

The decoupling given by equation (9) is just a specific case of a more general principle of decoupling, which will be described in the following. Let a function of many variables, $G_\Omega(x_1, x_2,...)$, be absolutely bounded and have the property that for large enough $\Omega$

$$|G_\Omega(x_1, x_2,...) - G_\Omega(0, x_2,...)| < g(\Omega) f(x_1, x_2,...), \tag{19}$$

where $g(\Omega)$ tends to zero as $\Omega$ tends to infinity and let the variables $\{x_i\}$ be random and independent, then,

$$\lim \Omega \to \infty \langle h(x_1) G_\Omega(x_1, x_2,...) \rangle = \langle h(x_1) \rangle \lim \Omega \to \infty \langle G_\Omega(x_1, x_2,...) \rangle, \tag{20}$$

which is very simple to understand.